\documentclass{mem}
\usepackage{natbib}
\usepackage{txfonts}
\usepackage{balance}
\usepackage{graphicx}
\usepackage{enumitem}
\usepackage{amssymb}
\usepackage{wrapfig}

\idline{75}{282}
\begin{document}
\def\teff{$T\rm_{eff }$}
\def\kms{$\mathrm {km s}^{-1}$}

\title{
The Soft X-ray Imager (SXI) on-board the THESEUS mission
}

   \subtitle{}

\author{
P. \,O'Brien\inst{1} 
\and
E. Bozzo\inst{2}
\and
R. Willingale\inst{1}
\and
I. Hutchinson\inst{1}
\and
J. Osborne\inst{1}
\and
L. Amati\inst{3}
\and
D. G\"otz\inst{4}
          }

\institute{
Department of Physics and Astronomy, 
University of Leicester, Leicester LE1 7RH, UK
\email{pto2@leicester.ac.uk}
\and
ISDC Data Centre for Astrophysics, Chemin d'Ecogia 16, 1290, Versoix, Switzerland
\and
LAM Laboratoire d'Astrophysique de Marseille, rue Frederic Joliot-Curie 38, F-13388
Marseille Cedex 13, France
\and
CEA Saclay - Irfu/D\'epartement d'Astrophysique, Orme des Merisiers, B\^at. 709, F-91191 Gif-sur-Yvette, France
}

\authorrunning{P. O'Brien}

\titlerunning{The THESEUS SXI}

\abstract{We summarize in this contribution the capabilities, design status, and the enabling technologies of the Soft X-ray Imager (SXI) 
planned to be on-board the THESEUS mission. We describe its central role in making THESEUS a powerful machine to probe the physical 
conditions of the early Universe (close to the reionization era) and to explore the time-domain Universe.

}
\maketitle{}

\section{Introduction}

The Transient High Energy Sky and Early Universe Surveyor (THESEUS) is a space mission concept 
developed in response to the fifth call for medium-sized missions by the European Space Agency (ESA).  
The core science goals of THESEUS comprise the exploration of the cosmic dawn and re-ionization era through the prompt detection and characterization of  
Gamma-Ray Bursts (GRBs) up to high redshift ($z\gtrsim$10), as well as a vast improvement of our knowledge of the soft X-ray transient Universe. 
THESEUS is also expected to critically complement the breath of multi-messenger facilities that will operate in the late 2020s. 
\begin{figure*}[t!]
\centering
\includegraphics[scale=0.8]{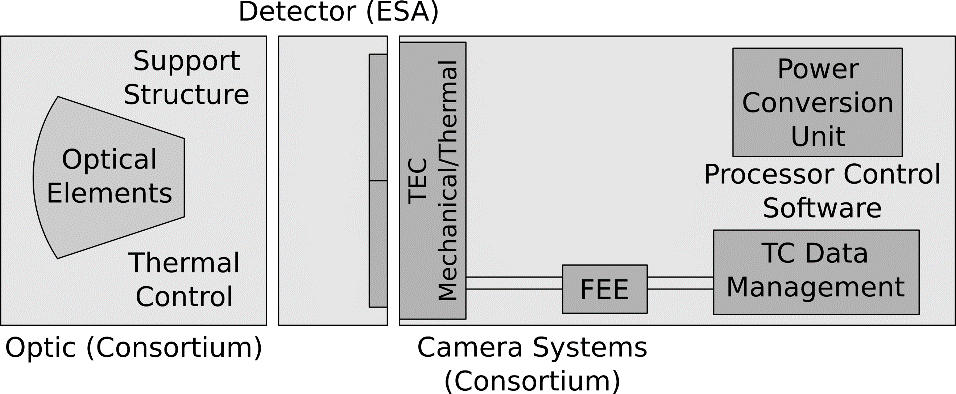}
\includegraphics[scale=0.8]{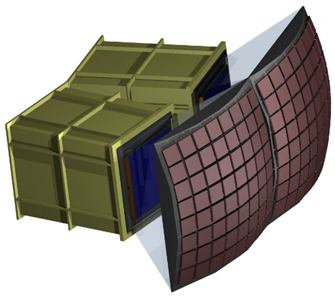}
\caption{The SXI block diagram concept (left) and the optical elements (right).}
\label{fig:26}   
\end{figure*}

In order to fulfill its science objectives, the THESEUS payload comprises: 
\begin{itemize}
\item A Soft X-ray Imager (SXI), featuring a set of 4 lobster-eye telescopes with a total field-of-view (FoV) of $\sim$1~sr. The SXI is capable to achieve a 
source locaization accuracy of $<$1-2~arcmin and operates in the 0.3-6~keV energy band. 

\item An InfraRed Telescope (IRT) with a 0.7~m mirror. IRT will provide a FoV of 10$\times$10~arcmin and the possibility to perform both 
imaging and spectroscopy observations in the 0.7-1.8~$\mu$m energy range. 

\item A X-Gamma ray Imaging Spectrometer (XGIS) made by a set of coded-mask cameras using monolithic X-gamma ray detectors. These are based on bars of Silicon 
diodes coupled with CsI crystal scintillator, granting a $\sim$1.5~sr FOV, a source location accuracy of $\sim$5~arcmin in 2-30~keV, and an unprecedently 
broad energy band (2~keV-20~MeV). 
\end{itemize}
\begin{table}
\centering
\scriptsize
\caption{Main properties of the THESEUS SXI DUs.}
\label{tab:4}  
\begin{tabular}{ll}
\hline\noalign{\smallskip}
Energy band (keV) & 0.3-6  \\
Telescope type & Lobster eye \\
Optics aperture & 320$\times$320~mm$^2$  \\
Optics configuration & 8$\times$8 square pore MCPs  \\
MCP size & 40$\times$40~mm$^2$ \\
Focal length & 300~mm \\
Focal plane shape & spherical \\
Focal plane detectors & CCD array \\
Size of each CCD & 81.2$\times$67.7~mm$^2$ \\
Pixel size & 18~$\mu$m \\
Number of pixel & 4510$\times$3758 per CCD \\
Number of CCDs & 4 \\
Field of View & $\sim$1~sr \\
Angular accuracy (best, worst) & ($<$10, 105)~arcsec\\
\noalign{\smallskip}\hline
\end{tabular}
\end{table}

The SXI and the XGIS will be able to detect GRBs and transient sources over large fractions of the sky.  
The SXI is expected to be the prime instrument for the detection of GRBs at high redshift, as its energy coverage is 
optimized to provide a high sensitivity in the soft X-ray domain. For many sources triggering the SXI,   
the XGIS will provide a refined spectral and timing characterization, allowing us to   
distinguish between GRBs and other classses of high energy transients. The IRT will take advantage of the 
fast slewing capability of the satellite (several degrees per minute) to provide a prompt identification of 
the IR counterpart of each triggering source. This will refine its position down to the $\sim$arcsec accuracy and (where relevant) 
determine the on-board redshift, providing also a spectroscopic characterization of both the counterpart and its host galaxy. 
The prompt availability of the refined source position will also enable follow-up observational campaigns with the largest ground and 
space observatories operating at the time of THESEUS. The unique combination of energy coverage, sensitivity, and fast repointing 
capabilities of THESEUS will guarantee an optimal synergy with the next-generation of gravitational 
wave/neutrino detectors, as well as the large electromagnetic (EM) facilities of the next decade \citep{stratta18}.  

In this contribution, we focus on the SXI, providing a description of the enabling technologies, the capabilities of the instrument and 
a description of the developing plan foreseen during the ESA M5 mission timeline. An exhaustive description of the THESEUS mission and 
the corresponding science objectives can be found in \citep{amati17}. 
\begin{figure*}[t!]
\centering
\includegraphics[scale=0.45]{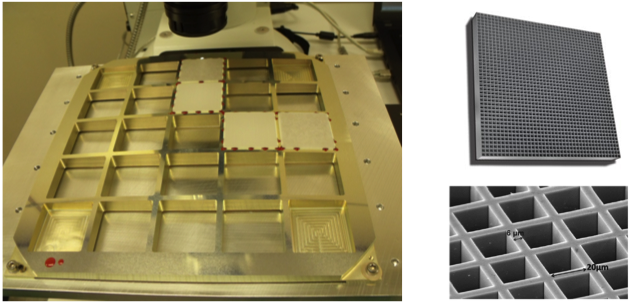}
\caption{{\it Left}: The SVOM MXT lobster eye optic aperture frame. {\it Top right}: A schematic of a single square pore MCP. {\it Bottom right}: 
A micrograph of a square pore MCP showing the pore structure. This plate has a pore size d=20~$\mu$m and a wall thickness w=6~$\mu$m.}
\label{fig:27}   
\end{figure*}

\section{The Soft X-ray Imager}

\begin{figure*}[t!]
\centering
\includegraphics[scale=0.16]{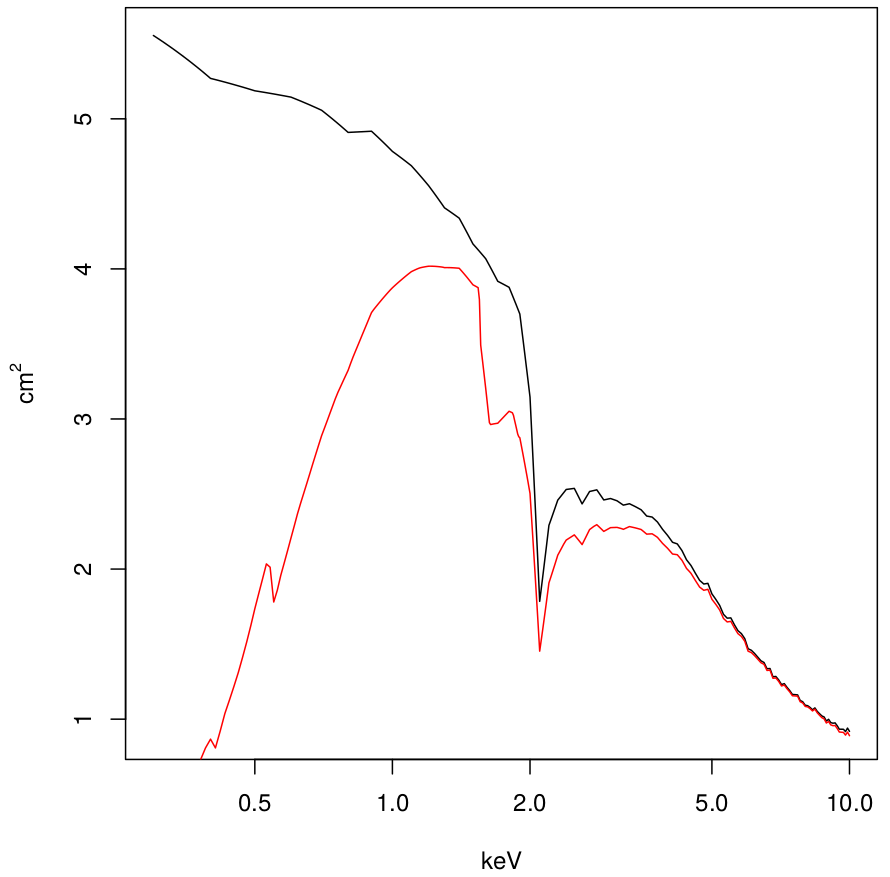}
\includegraphics[scale=1]{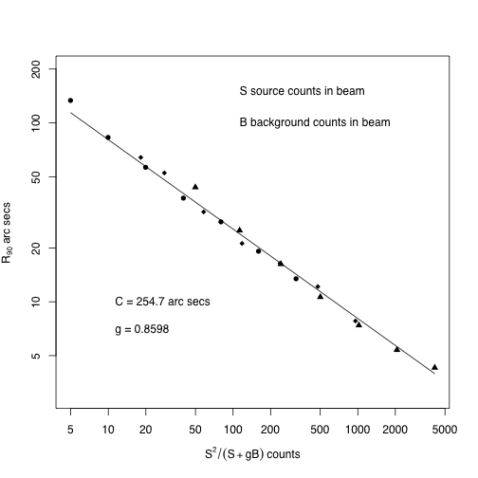}
\caption{{\it Left}: Collecting area as a function of energy (assuming a focal lenght of 300~mm and including the contributions 
of the central spot, the 2 cross-arms, and the straight through flux). The black line represents the optics only. 
The red curve includes the quantum efficiency of the CCD 
and the transmission of the optical blocking filter (comprising 60~nm Aluminium film deposited over the front of the MCPs and 260~nm of 
Aluminium plus 500~nm of parylene on the surface of the detectors). Because the angular width of the optics MCP-array 
is 2.3~deg larger than the CCD-array the field of view is unvignetted at 1~keV and above, so the collecting area 
shown here is constant across the field of view. {\it Right}: The position accuracy of the SXI as a function of source 
and background count. R90 is the error radius that contains 90\% of the derived positions.}
\label{fig:29}   
\end{figure*}

The SXI comprises four wide field lobster eye telescopes (each called a Detector Unit, DU), designed on 
the optical principle first described by \citet{Angel1979} and characterized by an optical bench as shown 
in Fig.~\ref{fig:26}. 
\begin{figure}
\centering
\includegraphics[scale=0.28]{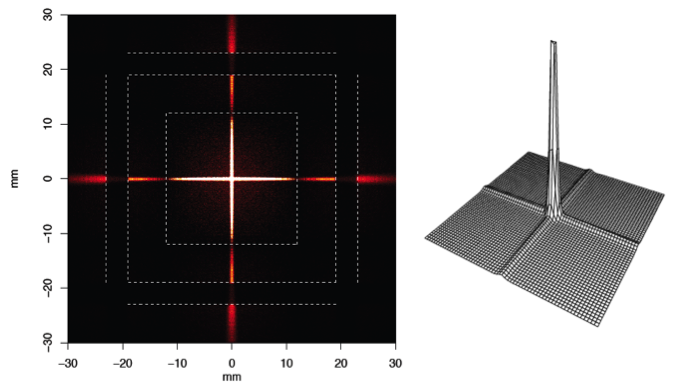}
\caption{The point spread function of the SXI.}
\label{fig:28}   
\end{figure}
The optics aperture is formed by an array of 8$\times$8 square pore Micro Channel Plates (MCPs), 
which are 40$\times$40~mm$^2$ wide and are mounted on a spherical frame with a radius of curvature of 
600~mm (2 times the focal length of 300~mm). The mechanical  
envelope of each SXI DU is characterized by a square cross-section of 320$\times$320~mm$^2$ at the optics end, 
tapering to 200$\times$200~mm$^2$ at detector. The depth of the detector housing is 200~mm, 
giving an overall module length of 500~mm.  
The left-hand side of Fig.~\ref{fig:27} shows the optics frame of the SVOM MXT lobster-eye telescope breadboard model, comprising  
21 square MCPs mounted over a 5$\times$5 grid. The front surface is spherical and its radius of curvature is of 2000~mm, resulting in a focal length of 1000~mm. 
The design proposed for the SXI uses the same plate size and exactly the same mounting principle but a shorter focal length (300~mm). 
The radius of curvature of the front surface is thus 600~mm. The right-hand panels of 
Fig.~\ref{fig:27} shows a schematic of a single plate and a micrograph that reveals the square pore glass structure. The focal plane of each SXI 
module is a spherical surface of radius of curvature 600~mm situated at a distance of 300~mm (the focal length) from the optics aperture. The 
detectors for each module comprise a 2$\times$2 array of large format detectors tilted to approximate to the spherical focal surface. 
The typical point-spread function (PSF) of an X-ray source observed by a lobster-eye telescope is shown in Fig.~\ref{fig:28}. 
The total collecting area of each DU is shown in Fig.~\ref{fig:29}. 

\begin{figure*}[t!]
\centering
\includegraphics[scale=0.2]{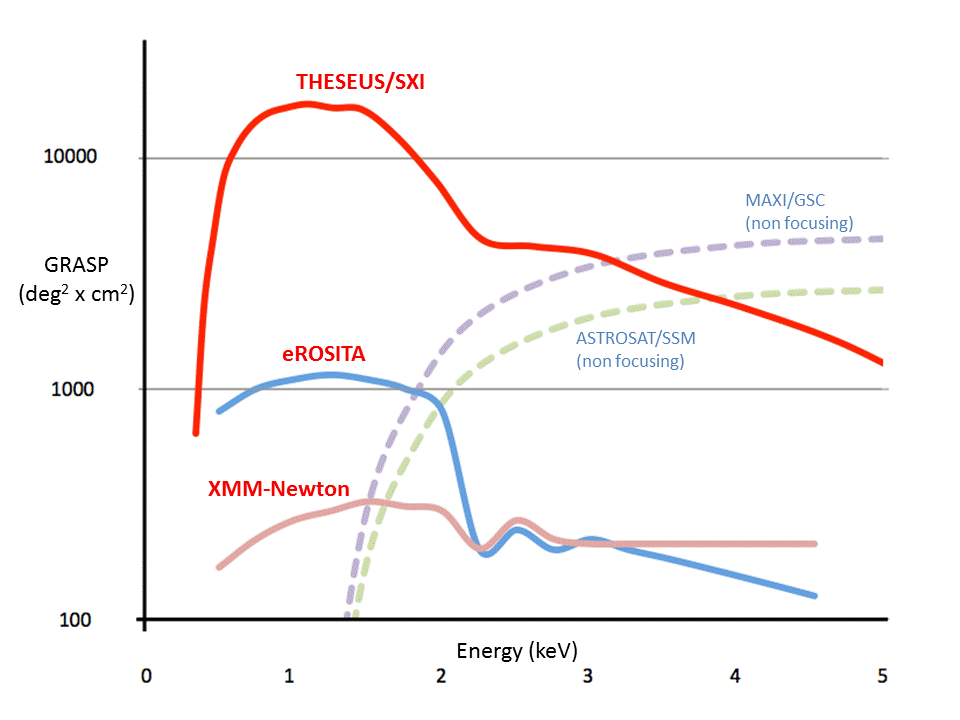}
\caption{The GRASP (FoV$\times$Effective Area) of the SXI in the soft X-ray energy band compared to XMM-Newton 
and eROSITA. The GRASP of X-ray monitors on-board MAXI and ASTROSAT are also show for completeness, even though these are not focusing 
and their sensitivity for a given effective area is substantially worse than that of focusing telescopes.}
\label{fig:23}   
\end{figure*}

The PSF is characterized by a 
prominent central spike, containing about 1/4 of the total source counts, and four widely extended cross-arms in which the remaining 3/4 of the counts 
are recorded. The inner dotted square in the left plot of Fig.~\ref{fig:28} shows the off-axis angle at which the count-rate in the 
cross arms goes to zero, as determined by the L/d ratio of the MCP pores. The value L/d=50 has been shown to optimize the performances 
at 1~keV. The outer dotted square indicates the shadowing of the cross-arms introduced by the gap between the individual MCPs in 
the aperture. The central true-focus spot is illustrated by the projection plot to the left. The FWHM is 4.5~arcmin  
and all the true-focus flux is contained by a circular beam of diameter 10~arcmin. The imaging area of the CCDs sets the field of 
view of each module and with four of them the total field of 
view would be of 3200~square degrees (0.9~sr). 
The characteristics of the SXI are summarized in Tab.~\ref{tab:4}. We also show a comparison of the SXI grasp 
(FoV$\times$Effective Area) compared to previous and currently operating facilities in Fig.~\ref{fig:23}.

\begin{figure*}[t!]
\centering
\includegraphics[scale=1.0]{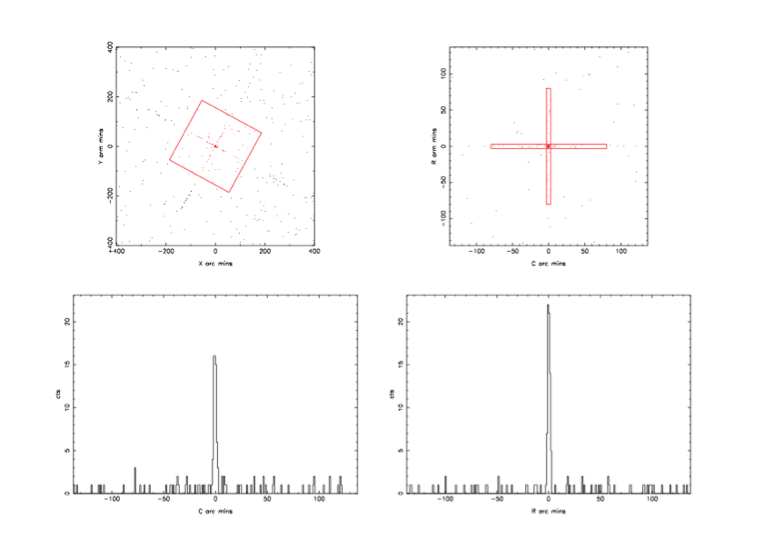}
\caption{The two stage trigger algorithm. {\it Top left-hand panel}: the detected event distribution $\Delta$T=4 seconds and a 
source count rate of 40~Cts/s over the full PSF. The cross-arms are rotated wrt the detector axes to demonstrate how this 
can be handled. {\it Top right-hand panel}: the detected event distribution in the patch of sky aligned to the cross-arm axes 
of the PSF (shown as the red rectangle in the top left-hand panel). The red cross-patch indicates the area used for the second stage of the algorithm. 
{\it Bottom panels}: the histograms along columns and rows in the patch.}
\label{fig:30}   
\end{figure*}

An X-ray source observed within the FoV of the SXI will trigger the instrument and its emission properties will be recorded 
through an optimized detection algorithm. The ideal algorithm would be some form of matched filtering using the full PSF distribution 
but because of the extent of the PSF this is too computationally heavy. A much simpler and computationally affordable scheme would be to 
search for significant peaks using the cell size commensurate with the central spike of the PSF. The disadvantage is that this method would 
only use $\sim$25\% of the total flux detected. The most suitable option considered so far is a two stage process 
which exploits the cross-arm geometry but avoids computationally expensive 2-D cross-correlation. This is summarized below and in 
Fig.~\ref{fig:30}. The focal plane is first divided into square patches with angular side lengths of  
$\sim4d/L=1/12$ radians, aligned with the cross-arm axes. The dotted central square shown in Fig.~\ref{fig:30} indicates the size and 
orientation of such a patch. The optimum size of the patches depends on the half energy width (HEW) of the lobster-eye optic 
and the background count rate. The patches could correspond to the detector elements or tiles in the focal plane, e.g. the CCD arrays. 
The peak profile is the line spread profile of the central spot and cross-arms of the PSF. The remainder of the histogram distribution arises 
from events in the cross-arm parallel to the histogram direction, the diffuse component of the PSF and any diffuse background 
events not associated with the source. Because we are looking for transient sources, the fixed pattern of the steady sources 
in the field of view at the time would have to be subtracted from the histogram distributions. As the pointing changes, the  
fixed pattern background would need to be updated. A transient source is detected if a significant peak is seen in both histograms.

The sensitivity of the detection and the accuracy of the derived source position within the patch depends on the bin size of 
the histograms, the HEW of the central peak of the PSF, and the background (see Fig.~\ref{fig:29}). For the most sensitive detection, the bin size 
should be approximately equal to the HEW but this will limit the accuracy of the position. If the bin size of the histograms 
is chosen to be significantly smaller than the width of the HEW, then the histograms can be smoothed by cross-correlation with 
the expected line width profile of the peak-cross-arm combination. Using the smaller bin size, the histograms can also be used 
to estimate the position centroid of the source within the patch. The significance used for this first stage should be low, 
e.g. 2.5~$\sigma$. This will provide candidate positions for the second stage.
For each of the candidate positions identified in the first stage, a cross-arm patch is set up to cover just the detector 
area which is expected to contain a fraction of the full cross-arms and the central peak in the PSF. The cross-patch dimensions 
are changed depending on the integration time $\Delta$T. For short integration times, the total background count will be small and 
the cross-patch size is set large to capture a large fraction of the counts from the cross-arms and the central peak. 

We envisage that a series of searches would be run in parallel, each using a 
different integration time so that the sensitivity limit as a function of $\Delta$T is covered. The basic scheme is illustrated 
in Fig.~\ref{fig:30}. The total source count assumed for this illustration was 40~Cts spread over the full PSF as plotted in Fig.~\ref{fig:28}. 
The bin size used for the histograms was 1~arcmin, and the HEW of the central peak of the PSF is approximately 4~arcmins. 
We have tested the algorithm over a range of integration times and background conditions. 
When a significant transient peak is identified the position must 
be converted to sky coordinates using the current aspect solution (from the on-board star-trackers). 
The position of any source which is identified as a possible interesting target (e.g., through cross-correlation of different 
catalogues) should be passed to the spacecraft for the automatic slewing.  
\begin{figure*}[t!]
\centering
\includegraphics[scale=0.5]{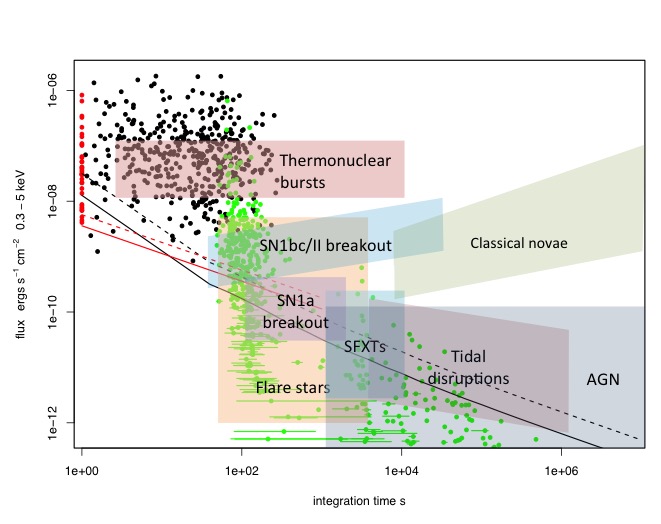}
\caption{Sensitivity of the SXI (black curves) and XGIS (red) vs. integration time. 
The solid curves assume a source column density 
of 5$\times$10$^{20}$~cm$^{-2}$ (i.e., well out of the Galactic plane and very little intrinsic absorption). The dotted curves assume a source column 
density of 10$^{22}$~cm$^{-2}$ (significant intrinsic absorption). The black dots are the peak fluxes for Swift BAT GRBs plotted against T90/2. 
The flux in the soft band 0.3-10~keV was estimated using the T90 BAT spectral fit including the absorption from the XRT spectral fit. 
The red dots are those GRBs for which T90/2 is less than 1~s. The green dots are the initial fluxes and times since trigger at 
the start of the Swift XRT GRB light-curves. The horizontal lines indicate the duration of the first time bin in the XRT light-curve. 
The various shaded regions illustrate variability and flux regions for different types of transients and variable sources.}
\label{fig:24}   
\end{figure*}

\subsection{Scientific performances}

The SXI design and capabilities have been optimized in order to allow THESEUS to achieve all its science goals. 
In particular, the SXI has a central role in the detection of GRBs at high redshifts, which are crucial for THESEUS 
to probe the physical conditions of the early Universe close to the epoch of the reionization. The large number of 
high redshift GRBs discovered with the SXI will improve our knowledge of the star formation rate, measure the 
high-z galaxy luminosity function, and demonstrate the role of Pop-III stars in the chemical enrichment 
of their environment. As detailed in \citet{amati17}, it is expected that THESEUS will detect with the SXI 
between 30 and 80 GRBs at z$>$6 over a three year-mission, with between 10 and 25 of these at z$>$8 (and sev-
eral at z$>$10). 

The SXI will also have a central role in the detection of electro-magnetic (EM) counterpart of neutrino and/or gravitational 
wave (GW) events. The combination of the instrument large FoV and sensitivity will be perfectly suited to cover the relatively 
large error boxes provided by the neutrino and GW facilities, detecting even possible faint X-ray counterparts. If compared to 
Swift/XRT, the improvement in the SXI grasp (FoV$\times$Effective area) is of a factor of $\sim$150. In case of a GW event from a 
neutron star-neutron star or neutron star-blackhole mergers, the faint X-ray afterglow of the associated short GRB (caught by the THESEUS/XGIS) 
can be detected by the SXI up to z$\gtrsim$2 or larger. 

The SXI will also observe a large number of different transient astrophysical sources, critically contributing to 
make THESEUS a powerful machine to explore the time-domain Universe. Using the Rosat All-sky Survey data we can 
estimate the count rate expected from the diffuse sky (Galactic and Cosmic) and point sources. 
The sensitivity to transient sources using this background rate and a false detection probability of 1.0$\times$10$^{-10}$ is shown as a 
function of integration time in Fig.~\ref{fig:24}. For longer integration times the source count required for a significant detection 
rises, e.g., to 30 counts for a 1~ks integration.


\bibliographystyle{aa}
\bibliography{sxi}

\begin{thebibliography}{3}
\expandafter\ifx\csname natexlab\endcsname\relax\def\natexlab#1{#1}\fi

\bibitem[{{Amati} {et~al.}(2017){Amati}, {O'Brien}, {Goetz}, {Bozzo}, {Tenzer},
  {Frontera}, {Ghirlanda}, {Labanti}, {Osborne}, {Stratta}, {Tanvir},
  {Willingale}, {Attina}, {Campana}, {Castro-Tirado}, {Contini}, {Fuschino},
  {Gomboc}, {Hudec}, {Orleanski}, {Renotte}, {Rodic}, {Bagoly}, {Blain},
  {Callanan}, {Covino}, {Ferrara}, {Le Floch}, {Marisaldi}, {Mereghetti},
  {Rosati}, {Vacchi}, {D'Avanzo}, {Giommi}, {Gomboc}, {Piranomonte}, {Piro},
  {Reglero}, {Rossi}, {Santangelo}, {Salvaterra}, {Tagliaferri}, {Vergani},
  {Vinciguerra}, {Briggs}, {Campolongo}, {Ciolfi}, {Connaughton}, {Cordier},
  {Morelli}, {Orlandini}, {Adami}, {Argan}, {Atteia}, {Auricchio}, {Balazs},
  {Baldazzi}, {Basa}, {Basak}, {Bellutti}, {Bernardini}, {Bertuccio}, {Braga},
  {Branchesi}, {Brandt}, {Brocato}, {Budtz-Jorgensen}, {Bulgarelli}, {Burderi},
  {Camp}, {Capozziello}, {Caruana}, {Casella}, {Cenko}, {Chardonnet}, {Ciardi},
  {Colafrancesco}, {Dainotti}, {D'Elia}, {De Martino}, {De Pasquale}, {Del
  Monte}, {Della Valle}, {Drago}, {Evangelista}, {Feroci}, {Finelli},
  {Fiorini}, {Fynbo}, {Gal-Yam}, {Gendre}, {Ghisellini}, {Grado}, {Guidorzi},
  {Hafizi}, {Hanlon}, {Hjorth}, {Izzo}, {Kiss}, {Kumar}, {Kuvvetli}, {Lavagna},
  {Li}, {Longo}, {Lyutikov}, {Maio}, {Maiorano}, {Malcovati}, {Malesani},
  {Margutti}, {Martin-Carrillo}, {Masetti}, {McBreen}, {Mignani}, {Morgante},
  {Mundell}, {Nargaard-Nielsen}, {Nicastro}, {Palazzi}, {Paltani}, {Panessa},
  {Pareschi}, {Pe'er}, {Penacchioni}, {Pian}, {Piedipalumbo}, {Piran}, {Rauw},
  {Razzano}, {Read}, {Rezzolla}, {Romano}, {Ruffini}, {Savaglio}, {Sguera},
  {Schady}, {Skidmore}, {Song}, {Stanway}, {Starling}, {Topinka}, {Troja}, {van
  Putten}, {Vanzella}, {Vercellone}, {Wilson-Hodge}, {Yonetoku}, {Zampa},
  {Zampa}, {Zhang}, {Zhang}, {Zhang}, {Zhang}, {Antonelli}, {Bianco}, {Boci},
  {Boer}, {Botticella}, {Boulade}, {Butler}, {Campana}, {Capitanio}, {Celotti},
  {Chen}, {Colpi}, {Comastri}, {Cuby}, {Dadina}, {De Luca}, {Dong}, {Ettori},
  {Gandhi}, {Geza}, {Greiner}, {Guiriec}, {Harms}, {Hernanz}, {Hornstrup},
  {Hutchinson}, {Israel}, {Jonker}, {Kaneko}, {Kawai}, {Wiersema}, {Korpela},
  {Lebrun}, {Lu}, {MacFadyen}, {Malaguti}, {Maraschi}, {Melandri}, {Modjaz},
  {Morris}, {Omodei}, {Paizis}, {Pata}, {Petrosian}, {Rachevski}, {Rhoads},
  {Ryde}, {Sabau-Graziati}, {Shigehiro}, {Sims}, {Soomin}, {Szecsi}, {Urata},
  {Uslenghi}, {Valenziano}, {Vianello}, {Vojtech}, {Watson}, \&
  {Zicha}}]{amati17}
{Amati}, L., {O'Brien}, P., {Goetz}, D., {et~al.} 2017, arXiv/1710.04638

\bibitem[{{Angel}(1979)}]{Angel1979}
{Angel}, J.~R.~P. 1979, \apj, 233, 364

\bibitem[{{Stratta} {et~al.}(2017){Stratta}, {Ciolfi}, {Amati}, {Ghirlanda},
  {Tanvir}, {Bozzo}, {Gotz}, {O'Brien}, {Frontera}, {Osborne}, {Rezzolla},
  {Rossi}, {Maiorano}, {Vinciguerra}, {Guidorzi}, {Drago}, {Nicastro},
  {Palazzi}, {Branchesi}, {Boer}, {Brocato}, {Bulgarelli}, {Covino}, {D'Elia},
  {Dainotti}, {De Pasquale}, {Gendre}, {Jonker}, {Longo}, {Mereghetti},
  {Mignani}, {Mundell}, {Piranomonte}, {Razzano}, {Sz{\'e}csi}, {van Putten},
  {Zhang}, {Hudec}, {Vergani}, {Malesani}, {D'Avanzo}, {Colafrancesco},
  {Stamerra}, {Caruana}, {Starling}, {Willingale}, {Salvaterra}, {Maio},
  {Greiner}, {Rosati}, {Labanti}, {Fuschino}, {Riccardo}, {Grado}, {Colpi},
  {Rodic}, {Patricelli}, \& {Bernardini}}]{stratta18}
{Stratta}, G., {Ciolfi}, R., {Amati}, L., {et~al.} 2017, arXiv/1712.08153

\end{thebibliography}

\end{document}